\documentclass[aps,prl,showpacs,twocolumn]{revtex4}
\usepackage{graphicx}
\input{epsf}

\begin{document}
\title {The origin of flux-flow resistance oscillations in
Bi$_2$Sr$_2$CaCu$_2$O$_{8+y}$:\\
Fiske steps in a single junction?}

\author{A. V. Ustinov}

\affiliation {Physikalisches Institut III, Universit\"at
Erlangen-N\"urnberg, D-91058, Erlangen, Germany}

\author{N. F. Pedersen}

\affiliation {Oersted-DTU, Section of Electric Power Engineering,
The Technical University of Denmark, DK-2800 Lyngby, Denmark}

\date{\today}
\begin{abstract}
We propose an alternative explanation to the oscillations of the
flux-flow resistance found in several previously published
experiments with Bi$_2$Sr$_2$CaCu$_2$O$_{8+y}$ stacks. It has been
argued by the previous authors that the period of the oscillations
corresponding to the field needed to add one vortex per two
intrinsic Josephson junctions is associated with a moving
triangular lattice of vortices (out-of-phase mode), while the
period corresponding to one vortex per one junction is due to the
square lattice (in-phase mode). In contrast, we show that both
type of oscillations may occur in a single-layer Josephson
junction and thus the above interpretation is inconsistent.
\end{abstract}


\maketitle

Recently, a lot of attention has been focused on the possible use
of single crystals of Bi$_2$Sr$_2$CaCu$_2$O$_{8+y}$ as generators
of electromagnetic radiation in the THz range. Promising
experiments in that direction were reported by Hirata et al.
\cite{Hirata-PRL-2002} and shortly thereafter by Kakeya et al.
\cite{Kakeya-2002-2004} and Hatano et al. \cite{Hatano-ASC-2004}.
These experiments all showed oscillations in the flux-flow voltage
and flux-flow resistance when a large magnetic field (of order
several Tesla) was applied parallel to the $ab$ plane in the
presence of a small bias current in the $c$--direction (of order a
few percent of the critical current). The observed flux-flow
voltage oscillations typically showed two different oscillation
periods. At the lowest magnetic fields the period was $\Delta
H_{\rm T}=\Phi_0/(2sL)$ corresponding to one extra flux quantum
per \emph{two} layers in the stack. Here $\Phi_0$ is the magnetic
flux quantum, $s$ and $L$ are the thickness and the length of the
junction, respectively. At higher magnetic fields there was a
transition to a period $\Delta H_{\rm S}=\Phi_0/(sL)\,$, i.e.
corresponding to an extra flux quantum in \emph{every} layer.

\begin{figure}
\includegraphics[width=3.4in]{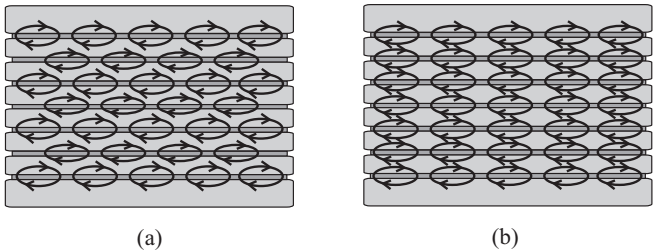}
\caption{Schematic drawing of the triangular lattice (a)
corresponding to the out-of-phase fluxon mode and the square
lattice (b) associated with the in-phase fluxon mode.}
\label{sketch-in-out-phase}
\end{figure}

These exiting experiments were interpreted both analytically
\cite{Koshelev-PRB-2002} and numerically
\cite{Machida-PRL-2003,Tachiki-2004,Peders-Madsen-IEEE-2005} by
several authors. One of the important questions to be answered was
whether the flux lattice correspond to a triangular lattice
(anti-phase ordering with possible cancellation of the sum voltage
at the junction end) or a square lattice (in-phase ordering
leading to a large sum voltage at the junction end). Figure
\ref{sketch-in-out-phase} shows schematically a
Bi$_2$Sr$_2$CaCu$_2$O$_{8+y}$ stack with flux ordering in a
triangular lattice and square lattice. Obviously, the latter case
is highly preferable for applications to THz generation of
electromagnetic waves; however simple intuition would suggest
triangular ordering since fluxons of same polarity naturally repel
each other. An intuitive interpretation of the experimentally
observed oscillations in the flux-flow voltage would suggest that
an oscillation period $\Delta H_{\rm T}$ corresponds to triangular
ordering while a period $\Delta H_{\rm S}$ corresponds to a square
lattice.  Numerical simulations
\cite{Machida-PRL-2003,Tachiki-2004,Peders-Madsen-IEEE-2005} have
shown that both triangular and square lattices are possible, but
their relation to regions of $\Delta H_{\rm T}$ oscillations and
regions of $\Delta H_{\rm S}$ oscillations is not simple and
details are still a matter of debate.

In this paper, we present fairly standard numerical simulations
corresponding to the simplest case of a single-layer Josephson
junction. Even for this case, where there is no triangular nor
square lattice ordering (as our stack consists of only one
junction), we find that both the $\Delta H_{\rm T}$ and $\Delta
H_{\rm S}$ periods appear much the same way as in experiments and
numerical simulations for Bi$_2$Sr$_2$CaCu$_2$O$_{8+y}$ stacks
with many layers and flux lattice ordering. After presenting the
numerical simulations we provide a qualitative explanation in
terms of the well-known Fiske modes \cite{Fiske-1964}.

The system under investigation is described by coupled sine-Gordon
equations of the form \cite{SBP-1993}:
\begin{equation}
\mathbf{S}\,\mathbf{J}=\frac{\partial^2 \varphi}{\partial x^2};
\;\;\; \mathbf{S} =
  \left(
    \begin{array}{cccccc}
      1   & S      &        &        &       &    \\
      S   & 1      & S      &        & 0     &    \\
          & \ddots & \ddots & \ddots &       &    \\
          &        & \ddots & \ddots & \ddots &   \\
          & 0      &        & S      & 1      & S \\
          &        &        &        & S      & 1
    \end{array}
  \right)
  \label{Eq:NPsG}
\end{equation}
with
\begin{equation}
J_i\equiv\frac{\partial^2 \varphi_i}{\partial
t^2}+\alpha\frac{\partial \varphi_i}{\partial
t}+\sin\varphi_i-\gamma
  \:\:.
  \label{J:NPsG}
\end{equation}
Here $\alpha=(1/R)[\hbar/(2eI_0C)]^{1/2}$ is the dissipation
parameter ($R$, $I_0$, and $C$ are the normal resistance, the
critical current and the capacitance per unit length,
respectively), $\gamma$ is the current normalized to the critical
current $I_0$ of the individual junctions. The normalized coupling
term among the junctions in the stack reads $S=-\lambda_{\rm
L}/[d'\sinh (t/\lambda_{\rm L})]$, where $d'=d+2\lambda_{\rm
L}\coth (t/\lambda_{\rm L})$, $t$ is the thickness of one
superconducting layer, and $\lambda_{\rm L}$ is the London
penetration depth \cite{SBP-1993}. Time $t$ is normalized to the
inverse of the Josephson plasma frequency $\omega_{\rm p}
=[2eI_0/(\hbar C_{\rm J})]^{1/2}$ and spatial coordinate $x$ is
normalized with respect to the Josephson length $\lambda_{\rm
J}=[\hbar/(2e\mu_0I_0)]^{1/2}$. The magnetic field gives rise to
boundary conditions of the form \cite{SBP-1993}
\begin{equation}
\left.\frac{\partial \varphi_i}{\partial x}\right|_{x=0,\ell}=
\frac{H}{\lambda_{\rm J}I_0} \equiv h\:.
  \label{bc:sG}
\end{equation}
Index $i=1,\ldots, N$ stands here for the junction number in the
stack and $\ell$ is the normalized length of the system.

\begin{figure}
\includegraphics[width=3.2in]{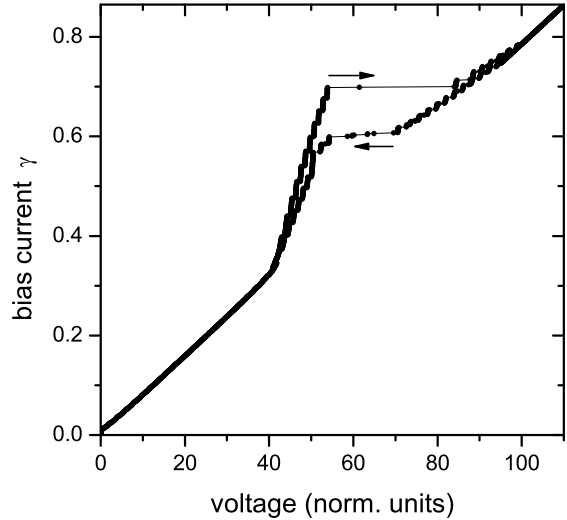}
\caption{Current-voltage characteristics of a long junction
($N=1$) with parameters $\ell=40$, $\alpha = 0.1$, and $h=4$.
Arrows indicate switching between branches for rising and
decreasing bias current $\gamma$.} \label{IV-curve}
\end{figure}

Figure \ref{IV-curve} shows the result of a simulation of the
current-voltage characteristics for only one junction in the
stack, i.e. $N=1$. It shows the flux-flow branch of the junction
with normalized length $\ell=40$ and damping constant $\alpha=0.1$
placed in magnetic field $h=4$. The characteristics is calculated
by rising the bias current $\gamma$ from zero to $0.85$ and then
decreasing it back to zero. The voltage $V$ is given in normalized
units chosen such that the voltage spacing between neighboring
Fiske steps
\begin{equation}
\Delta V=\frac{\Phi_0\bar c}{2L}\: \label{Fiske-volt-spacing}
\end{equation}
is equal to unity (here $\bar c$ is the Swihart velocity and
$L=\ell\lambda_{\rm J}$ is the physical length of the junction).
The current-voltage characteristics displays fine structure due to
the Fiske steps, which gather around the flux-flow voltage, also
known as Eck peak \cite{Eck-peak}. The hysteresis in this voltage
region is due to the coexistence of several Fiske resonances at a
given current $\gamma$. Some parts of these Fiske resonances are
located inside the hysteresis and are not displayed on the plot.

\begin{figure}
\includegraphics[width=3.2in]{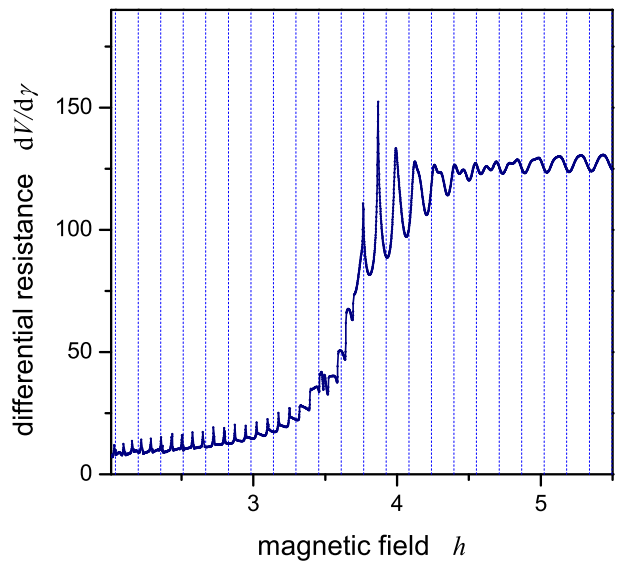}
\caption{Differential flux-flow resistance ${\rm d}V/{\rm
d}\gamma$ versus magnetic field $h$ for the single junction
($N=1$) with $\ell=40$ and $\alpha=0.1$ at constant bias current
$\gamma=0.3$.} \label{dV-dI-H}
\end{figure}
Using the above junction parameters, we calculated the dependence
of the flux-flow resistance on magnetic field. Figure
\ref{dV-dI-H} presents the differential flux-flow resistance ${\rm
d}V/{\rm d}\gamma$ versus magnetic field $h$ at the fixed bias
current $\gamma=0.3$. It clearly shows flux-flow resistance
oscillations. In order to relate the period of oscillations with
the number of vortices in the junction we show a grid in magnetic
field with a period $\Delta h=2\pi/L\ell$, which approximately
corresponds to adding one vortex in the junction. This can be seen
from the simple fact that at the critical field $h=2$ the
normalized spacing between vortices penetrated into the junction
is equal to $\pi$, and their number rises proportionally to $h$.
In Fig.~\ref{dV-dI-H} we see that most oscillations have a
characteristic period in $h$ corresponding to adding a half flux
quantum into the junction, with a tendency of doubling the period
at higher fields.

\begin{figure}
\includegraphics[width=3.2in]{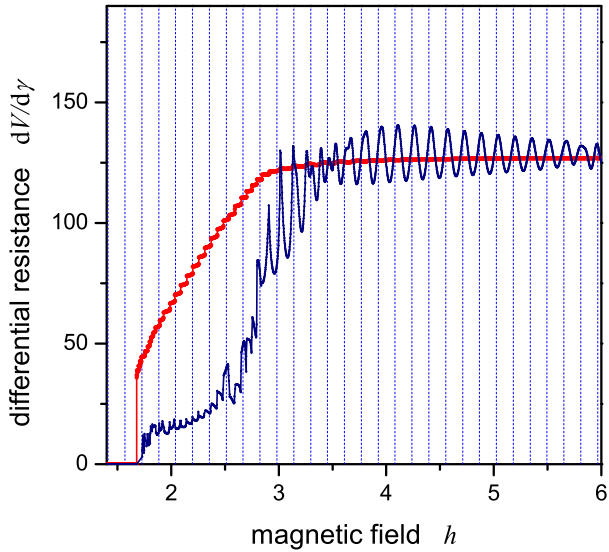}
\caption{Flux-flow resistance $V/\gamma$ (thick curve) and ${\rm
d}V/{\rm d}\gamma$ (thin curve) versus magnetic field $h$ for the
same junction as in Fig.~\protect\ref{dV-dI-H} at a lower bias
current $\gamma=0.2$.} \label{V-I-H}
\end{figure}
As another illustration, in Fig.~\ref{V-I-H} we show both static
resistance $V/\gamma$ and differential resistance ${\rm d}V/{\rm
d}\gamma$ versus magnetic field $h$ for the same junction but at a
lower bias current $\gamma=0.2$. The oscillations of ${\rm
d}V/{\rm d}\gamma$ (thin curve) have two characteristic periods,
which are found at different ranges of magnetic field $h$. The
oscillation period at low fields corresponds to about or less than
\emph{half flux quantum}, while at higher fields we find very
clear oscillations which account for \emph{one flux quantum} into
the junction. The crossover from one regime to another occurs at
magnetic field $h\approx 3$. Below this field the differential
resistance at $\gamma=0.2$ is lower as it is determined by the Eck
peak (see also Fig.~\ref{IV-curve}) composed of individual Fiske
steps. At $h>3$ the Eck peak shifts to higher currents and the
differential resistance levels at the resistive slope determined
by the loss parameter $\alpha$ of the junction. The static
junction resistance (upper curve in Fig.~\ref{V-I-H}) also changes
in this range but its oscillations are much less pronounced and
can only be clearly seen on the magnified scale.

We suggest the following explanation for the two oscillation
periods of the flux-flow resistance in a single-barrier Josephson
junction. At high enough magnetic field the resistance at a low
bias current follows oscillations of the critical current of the
junction and thus have a characteristic period of \emph{one flux
quantum}. At the same time, the resistance measured at a higher
current (or lower field but the same bias current) follows the
oscillations due to the Fiske steps, which envelope is associated
with the Eck peak (sometimes called as flux-flow or velocity
matching step). The matter is that -- for a single junction -- the
Fiske steps induce a variation of the resistance with
characteristic period corresponding to \emph{half flux quantum}.

Fiske modes in a Josephson junction \cite{Fiske-1964} are linear
cavity type excitations with resonance angular frequencies given
by
\begin{equation}
\omega_n=n\frac{2\pi}{\ell}\:, \:\:\:\: n=1,2,3,\ldots
  \label{Fiske-freq}
\end{equation}
in normalized units. The corresponding to wave vectors are
\begin{equation}
k_n=n\frac{\pi}{\ell}\:. \label{Fiske-vect}
\end{equation}
In experiments these Fiske modes are visible as current
singularities in the current-voltage curve with a voltage spacing
given by Eq.~(\ref{Fiske-volt-spacing}). The amplitude of the
Fiske steps oscillate with the magnetic field in a typical Bessel
function like pattern such that the even numbered steps have
maxima together with maxima in the critical current, while maxima
in the odd numbered steps correspond to minima in the critical
current. Refs.~\cite{Kulik,Matteo-1998} give an approximate
analytical form for the current-voltage curve for a single
Josephson junction with Fiske steps. The current-voltage
characteristics is approximately written as \cite{Matteo-1998}:
\begin{widetext}
\begin{equation}
\gamma(\omega,h)=\alpha\omega+ \sum_{-\infty}^{+\infty}\left[
\frac{h}{\ell \left({h^2-k_n^2}\right)} \sin\frac{h
\ell-k_n\ell}{2}\right]^2\cdot\frac{2\alpha
\omega}{\left(\omega^2-k_n^2\right)^2 + \alpha^2\omega^2}\:.
\label{Fiske-current}
\end{equation}
\end{widetext}
This approximation neglects high-order nonlinearities in the
junction and is originally expected \cite{Kulik} to describe well
the case of short junction, i.e. $\ell<1$. However, the comparison
with long junction data made by Cirillo et al. \cite{Matteo-1998}
shows that Eq.~(\ref{Fiske-current}) can approximately account for
the shape and for the maximum current modulation of the Fiske
singularities in long ($\ell\gg 1$) junctions when the field
penetration overcomes Meissner shielding, i.e. at $h>2$.

The first term in Eq.~(\ref{Fiske-current}) represents the Ohmic
part of the current-voltage characteristics, while the second term
gives an infinite series of equidistant resonances. The height of
the resonances is modulated by a slowly varying amplitude factor
and a fast Fraunhofer amplitude factor \cite{Matteo-1998}. The
Fraunhofer factor emphasizes the resonance closest to $\omega=h$
and drops off fast: If $h \ell$ is an even multiple of $\pi$, the
odd numbered Fiske steps are enhanced and if $h \ell$ is an odd
multiple of $\pi$, the even numbered Fiske steps survive.

\begin{figure}
\includegraphics[width=3.2in]{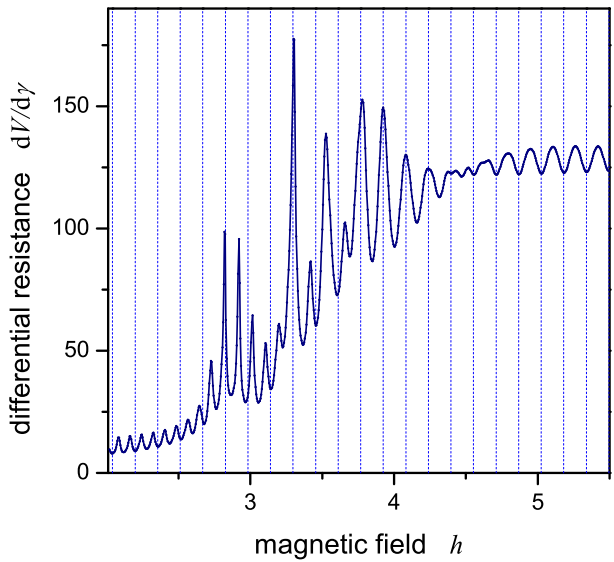}
\caption{Differential flux-flow resistance ${\rm d}V/{\rm
d}\gamma$ at constant bias current $\gamma=0.2$ versus magnetic
field $h$ calculated directly from
Eq.~(\protect\ref{Fiske-current}) with parameters $\ell=40$ and
$\alpha=0.1$. The voltage is multiplied by a factor $\ell/\pi$ to
make its scale identical to the voltage normalization used in the
previous figures.} \label{dV-dI-H-analyt}
\end{figure}
Equation~(\ref{Fiske-current}) gives the current-voltage curve
containing Fiske steps with the magnetic field as a parameter. If
we instead assume a fixed bias current $\gamma$ and vary the
magnetic field, Eq.~(\ref{Fiske-current}) expresses the voltage
$V\propto\omega$ oscillations in an implicit form. In order to
compare this analytical form with our simulations, we solved
Eq.~(\ref{Fiske-current}) for $\omega$ at a given $\gamma$
numerically. The obtained the dependence of the flux-flow
resistance ${\rm d}V/{\rm d}\gamma$ at constant bias current
$\gamma=0.2$ versus magnetic field $h$ is presented in
Fig.~\ref{dV-dI-H-analyt}. The qualitative agreement between
Figs.~\ref{dV-dI-H} and \ref{V-I-H} obtained by full numerical
simulation of the perturbed sine-Gordon equation and
Fig.~\ref{dV-dI-H-analyt} emerging from the analytical formula
(\ref{Fiske-current}) is strikingly good.
Figure~\ref{dV-dI-H-analyt} clearly displays two characteristic
periods of oscillations, namely half flux quantum oscillations at
low fields and one flux quantum oscillations which become very
explicit at high fields. The intermediate field range shows a
complicated beating between two periods.

Thus, odd and even numbered Fiske resonances in the last term in
Eq.~(\ref{Fiske-current}) produce the "magic" half-flux-quantum
oscillations corresponding to the magnetic field period $\Delta
H_{\rm T}=\Phi_0/(2sL)$ even in a single Josephson junction.
Although we investigated here Fiske steps with $N=1$, we note that
Fiske steps are also present in stacks with $N>1$
\cite{Kim-Hatano:Plasma-2004}. Thus we are lead to suggest that
also for $N>1$ the flux-flow voltage oscillations have their
origin in the Fiske mode excitations.

We conclude that for Bi$_2$Sr$_2$CaCu$_2$O$_{8+y}$ stacks the
flux-flow voltage oscillations with two different periods in a
magnetic field have their origin in the Fiske mode excitations.
Thus the flux lattice ordering in either triangular or square
lattice is not directly related to the two periods of the
oscillations. We speculate that Fiske modes also existing in
stacks indirectly play a role for the flux lattice formation.

We would like to acknowledge discussions with T. Hatano, S. Kim,
A. E. Koshelev, S. Madsen, M. R. Samuelsen, H. B. Wang, and T.
Yamashita.


\begin{thebibliography}{99}

\bibitem{Hirata-PRL-2002} S. Ooi, T. Mochiku, and K. Hirata,
Phys. Rev. Lett. \textbf{89}, 247002 (2002)

\bibitem{Kakeya-2002-2004} I. Kakeya, M. Iwase, T. Yamamoto,
and K. Kadowaki, cond-mat/0503498 (2005); I. Kakeya et al. Physica
C \textbf{378},406 (2002)

\bibitem{Hatano-ASC-2004} T. Hatano et al., IEEE Trans Appl.
Superconduc., to be published

\bibitem{Koshelev-PRB-2002} A.E. Koshelev, Phys. Rev B
\textbf{66}, 224514 (2002)

\bibitem{Machida-PRL-2003} M. Machida, Phys. Rev. Lett. \textbf{90},
037001 (2003)

\bibitem{Tachiki-2004} M. Tachiki, M. Iizuka, K. Minami,
S. Tejima, and H. Nakamura, cond-mat/0407052 (2004)

\bibitem{Peders-Madsen-IEEE-2005} N.F. Pedersen and S. Madsen,
IEEE Trans Appl. Superconduc., to be published

\bibitem{Fiske-1964} M. D. Fiske, Rev. Mod. Phys. \textbf{36},
221 (1964)

\bibitem{SBP-1993} S. Sakai, P. Bodin, and N.F. Pedersen,
J. Appl. Phys. \textbf{73}, 2411 (1993)

\bibitem{Eck-peak} R. E. Eck, D. J. Scalapino and B. N. Taylor,
Phys. Rev. Lett. \textbf{13}, 15 (1964).

\bibitem{Kulik} I. O. Kulik, Pis'ma Zh. Eksp. Teor.
Fiz. {\bf 2}, 134 (1965) [Sov. Phys. JETP Lett. {\bf 2}, 84
(1965)]; I. O. Kulik, Zh. Tekhn. Fiz. \textbf{37} 157 (1967) [Sov.
Tech. Phys. \textbf{12}, 111 (1967)]

\bibitem{Matteo-1998}  M. Cirillo, N. Gr{\o}nbech-Jensen, M. R.
Samuelsen, M. Salerno, and G. V. Rinati, Phys. Rev. B \textbf{58},
12377 (1998)

\bibitem{Kim-Hatano:Plasma-2004} S. Kim, S. Urayama,
H. B. Wang, T. Hatano, M. Nagao, S. Kawakami, Y. Takano, and T.
Yamashita. Poster Presentation in the conference 'Plasma-2004'
Tsukuba, Japan, November 2004



\end{thebibliography}
\end{document}